\documentclass[superscriptaddress,twocolumn]{revtex4}
\usepackage{amsmath,graphicx}

\begin{document}

\title {Joint measurement of current-phase relations and transport properties of hybrid junctions using a three junctions SQUID}
\author{J. Basset}
\affiliation{Laboratoire de Physique des Solides, Universit\'e Paris-Sud, CNRS, UMR 8502, F-91405 Orsay Cedex, France.}
\author{R. Delagrange}
\affiliation{Laboratoire de Physique des Solides, Universit\'e Paris-Sud, CNRS, UMR 8502, F-91405 Orsay Cedex, France.}
\author{R. Weil}
\affiliation{Laboratoire de Physique des Solides, Universit\'e Paris-Sud, CNRS, UMR 8502, F-91405 Orsay Cedex, France.}
\author{A. Kasumov}
\affiliation{Laboratoire de Physique des Solides, Universit\'e Paris-Sud, CNRS, UMR 8502, F-91405 Orsay Cedex, France.}
\author{H. Bouchiat}
\affiliation{Laboratoire de Physique des Solides, Universit\'e Paris-Sud, CNRS, UMR 8502, F-91405 Orsay Cedex, France.}
\author{R. Deblock}
\affiliation{Laboratoire de Physique des Solides, Universit\'e Paris-Sud, CNRS, UMR 8502, F-91405 Orsay Cedex, France.}

\begin{abstract}
We propose a scheme to measure both the current-phase relation and differential conductance $dI/dV$ of a superconducting junction, in the normal and the superconducting states. This is done using a dc Superconducting Quantum Interference Device (dc SQUID) with two Josephson junctions in parallel with the device under investigation and three contacts. As a demonstration we measure the current-phase relation and $dI/dV$ of a small Josephson junction and a carbon nanotube junction. In this latter case, in a regime where the nanotube is well conducting, we show that the non-sinusoidal current phase relation we find is consistent with the theory for a weak link, using the transmission extracted from the differential conductance in the normal state. This method holds great promise for future investigations of the current-phase relation of more exotic junctions.
\end{abstract} 

\maketitle 

\section{Introduction}

The prediction that a dissipationless current should flow between two superconductors separated by a thin insulator barrier was made in 1962 by Josephson \cite{josephson62}. Since then many works extended the validity of this prediction to other kinds of weak links such as narrow constrictions of the superconductor thin film or superconductor - normal metal - superconductor junctions (SNS)\cite{tinkham96,barone82,likharev79,golubov04,agrait03}. The supercurrent is expected to vary periodically with the phase difference between the two superconducting electrodes, with the maximum of this supercurrent being called the critical current of the junction. Probing the current phase relation (CPR) of the junction implies to phase bias it \cite{golubov04,agrait03,dellarocca07}. This is often done by inserting the weak link into a loop through which a magnetic flux is applied, thus constituting a SQUID. In the ac SQUID configuration only one junction is present in the loop whereas in the dc SQUID two junctions are present. In such configurations, the weak link is either short circuited (ac SQUID) or placed in parallel with another junction (dc SQUID) so that no direct conductance measurement can be made.  

In this work we propose a scheme to measure both the current phase relation and the differential conductance on the same sample. When the system is in the normal state, one wants to be able to measure the differential conductance of the weak link. This is needed to properly characterize the weak link. It has also a practical importance for carbon nanotube based weak links: it helps select the interesting devices \textit{i.e.} with relatively high conductance at room temperature. In the superconducting state, the weak link should be phase biased in order to measure the current-phase relation. We also want to be able to measure the differential conductance $dI/dV(V)$ of the weak link and thus voltage biased it. To reconcile these a priori contradictory requirements we propose to modify the dc SQUID geometry by introducing two reference Josephson junctions instead of one in the reference branch of the SQUID  and the weak link of interest in the other branch. A lead between the two reference Josephson junctions provides the extra probe necessary to characterize the weak link.    

The paper is organized as follows. In section \ref{sec:CPRsquid} we consider the influence of this geometry on the measurement of the CPR. We then present experimental results obtained when the weak-link is a small Josephson junction (section \ref{sec:CPR_JJ}) or a well conducting carbon nanotube (section \ref{sec:CPR_NT}).


\section{Current phase relation measurement with a SQUID geometry}
\label{sec:CPRsquid}

The aim of this work is to determine the CPR of a weak link, for example a S/carbon nanotube/S junction, with the possibility to also measure its differential conductance $dI/dV$ in the normal and superconducting states. In order to measure the CPR of the weak-link one needs to control its phase difference \textit{i.e.} realize a phase bias, which is also needed to probe the phase dependent Andreev bound state of the  weak link \cite{pillet10,chang13}. The phase bias for measuring the CPR of the weak link can be done by inserting it in a superconducting loop, in a dc SQUID geometry where the weak-link is in parallel with a Josephson junction. To measure reliably the CPR of the weak link one needs to consider a asymmetric SQUID where the supercurrent of the Josephson junction is much higher than the one of the weak link \cite{dellarocca07}. Taking into account the asymmetry of the junctions is generally an important point to understand the behaviour of DC SQUIDs \cite{kleiner96,greenberg03,novikov09}. In this geometry to unambiguously measure the conductance of one branch independently of the other, one technique consists in measuring the conductance of the SQUID, then opening one branch of the SQUID and measuring the conductance of the remaining branch and by subtraction, extract the conductance of the opened arm. This technique is used for break junctions \cite{dellarocca07}, high Tc junctions \cite{novikov09} and can be adapted to a gate tunable system with a strong reduction of the conductance \cite{cleuziou06,vanDam06,maurand12,franceschi10}. A second technique is to use the hysteretic $I(V)$ characteristics of the Josephson junction to obtain information on the small weak link. This however does not allow to extract the low voltage bias behaviour due to the finite bias voltage at which the phase is retrapped.
\begin{figure}[htbp]
  \begin{center}
			\includegraphics[width=8cm]{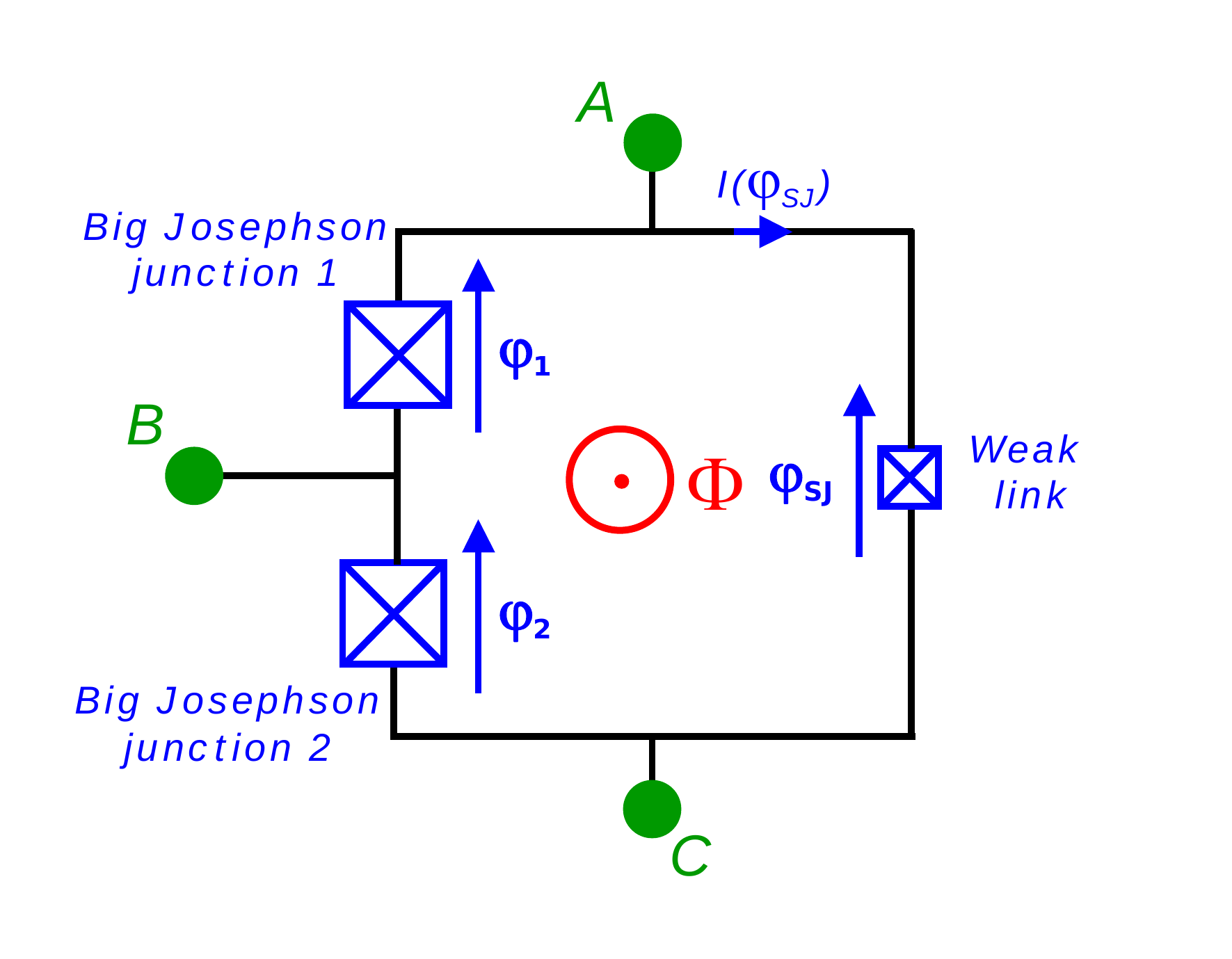}
	\end{center}
  \caption{Schematic picture of the dc SQUID with three contacts. Each junction has a superconducting phase difference $\varphi_1$, $\varphi_2$ and $\varphi_{SJ}$. Contacts $A$ and $C$ are respectively the SQUID source and drain. Contact $B$ is an additional central contact between the two almost identical big junctions.}
  \label{IdeaSquid}
\end{figure}

These difficulties can be overcome by the use of a three contacts SQUID geometry (Figure \ref{IdeaSquid}). One branch contains the weak link (small Josephson junction, carbon nanotube quantum dots,...) of interest. The other reference branch contains two nominally identical large Josephson junctions with a central contact (depicted by the letter $B$). The interest of this device is threefold. First the current phase relation of the weak link can be extracted from the modulation with magnetic flux of the asymmetric SQUID's switching current. Second the contact between the two large junctions enables the determination of the weak link's normal state resistance at room temperature. This is particularly useful to select hybrid junctions with low enough resistance, and more generally to check the quality of connection to the weak link. Third, one can use this same contact to measure the differential conductance of the weak link in the superconducting state without relying on the hysteresis loop of the junctions in parallel with the weak link.

In the following we will described the difference between the three junctions SQUIDs and the two junctions SQUID. 

\subsection{Resistively Capacitively Shunted Junction (RCSJ) model}

The current at which a Josephson junction switches from the zero resistance state to the resistive state (switching current $I_S$) is always smaller than the critical current $I_C$. This is related to the dynamics of the superconducting phase across the junction and has been widely studied in the last decades \cite{barone82,tinkham96}. In particular it was shown that the switching current depends on the electromagnetic environment in which the junction is embedded and can be described within the $RCSJ$ model. The junction is then considered as a perfect Josephson element in parallel with a $RC$ circuit. The superconducting phase difference across the junction is noted $\varphi$.

When the junction is current biased with a current $I=s I_C$, with $I_C$ the critical current of the junction (determined by theory \cite{ambegaokar63}) and $s\in[-1,1]$ a real number, one has :
\begin{equation}
  		I = I_{C} \sin (\varphi)+\frac{V}{R}+C\frac{dV}{dt}.
\end{equation}
Using the Josephson relation $2e V=\hbar d\varphi/dt$ one gets \cite{tinkham96}~: 
\begin{equation}
  		\phi_0 C \frac{d^2\varphi}{dt^2}+\frac{\phi_0}{R} \frac{d\varphi}{dt}+ I_C \sin (\varphi)=I
\label{equadiffRCSJ}  		
\end{equation}
with $\phi_0=\hbar/2e$. It is analog to the equation of motion of a particle of mass $m=\phi_0^2 C$ moving along the $\varphi$ axis in the effective washboard potential $U(\varphi)$ :
\begin{equation}
U(s,\varphi)=-\phi_0 I_C \cos (\varphi)-\phi_0 I \varphi,
\end{equation}
with a dissipation related to the quality factor $Q=\omega_P R C$ of the system, with $\omega_P=\sqrt{2e I_C/\hbar C}$ the plasma frequency of the junction. When the applied current is below the critical current, \textit{i.e.} $s<1$, the fictitious particle is trapped into a local minimum of the potential where it oscillates at $\omega_P(s)$. These oscillations are damped with a time scale $Q/\omega_P$. However, due to fluctuations in the current $\delta I(t)$, the particle can escape this local minimum leading to a finite dc voltage drop across the junction. Depending on the nature of the escape process, thermal or quantum, the escape rate has different expressions.
\begin{figure}[htbp]
  \begin{center}
			\includegraphics[width=8cm]{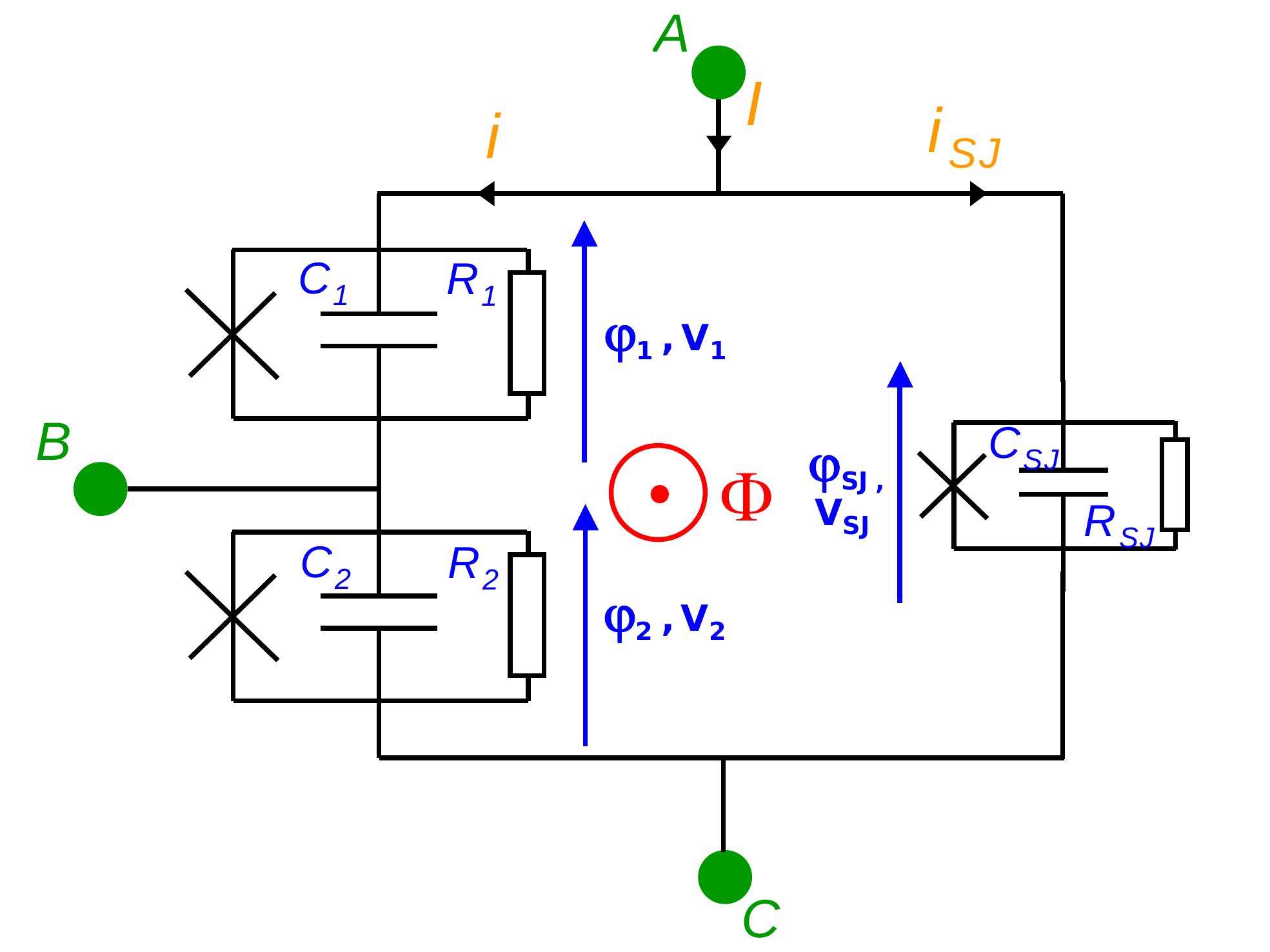}
	\end{center}
  \caption{RCSJ model for the asymmetric SQUID considered in the experiment.}
  \label{fig:RCSJSquid}
\end{figure}

\paragraph{Thermal escape}
The probability for the fictitious particle to escape from the well at a given current is given by $P_t(s)=1-e^{-\Gamma(s)t}$, with :
\begin{equation}
\Gamma(s)=a(Q)\frac{\omega_P(s)}{2\pi}e^{-\Delta U(s)/k_B T_{esc}}.
\label{rateswitch}
\end{equation}
$\omega_P(s)$ is the plasma frequency for current $s I_C$, the function $a(Q)$ accounts for friction and depends on the regime of the junction (Table \ref{tableAdeQ}) \cite{garg95,ruggiero98,petkovicPhD}, $T_{esc}$ corresponds to the escape temperature (temperature of the electromagnetic environment) and $\Delta U(s)$ is the height of the potential barrier at $I=s I_C$. 

\begin{table}[htbp]
\begin{center}
\begin{tabular}{|l|c|r|}
  \hline
  Damping & Validity range & a(Q) \\
  \hline
  \hline
   Underdamped, low & $Q>1$, $\ \frac{2 \pi \Delta U}{k_BT}\frac{\omega_0}{Q\omega_P}<1$ & $2\pi\frac{\Delta U}{k_BT}\frac{\omega_0}{Q\omega_P}<1$ \\
   \hline
   Underdamped, mod & $Q>1$, $\ \frac{2 \pi \Delta U}{k_BT}\frac{\omega_0}{Q\omega_P}>1$ & $1$ \\
   \hline
   Overdamped & $Q<1$ & $\frac{Q\omega_P}{\omega_0}$ \\
  \hline
\end{tabular}
\end{center}
\caption{Criterion for crossover between different damping regimes and the prefactor $a(Q)$ of the tunneling rate formula \ref{rateswitch}. In these formulas $\omega_0=\omega_P(s=0)$}  
\label{tableAdeQ}
\end{table}

\paragraph{Quantum escape}
In addition to thermal escape one has to consider quantum escape. Indeed since the phase is a quantum variable it may escape the potential well via quantum tunneling. The tunneling rate is well approximated in the underdamped regime by \cite{devoret85,martinis87}~:
\begin{equation}
\Gamma_{Tunnel}(s)=6^{3/2} \sqrt{\pi} \omega_P(s) \sqrt{\frac{\Delta U(s)}{\hbar\omega_P(s)}}e^{\frac{-36\Delta U(s)}{5\hbar\omega_P(s)}}.
\label{rateswitch2}
\end{equation}    
The crossover temperature between thermal and quantum escape is given by $T_{cross}=\hbar \omega_P(s=0) /2 \pi k_B$. 

In the following we will concentrate on the thermally activated behaviour, which is relevant for our experiments.

\subsection{Weak link embedded in a SQUID}
With a junction in parallel with a the Josephson junction, the previous model is modified. The capacitance C and conductance 1/R are now respectively the sum of the capacitances and the conductances of the Josephson junction and the probed junction. In this geometry, due to the fluxoid quantization, the phase difference of the small junction $\varphi_{SJ}$ is related to the phase difference $\varphi$ across the Josephson junction by~:
\begin{equation}
	\varphi_{SJ}=\varphi-2\pi\phi/\phi_0+2 n \pi
\end{equation}
where $\phi$ is the magnetic flux through the SQUID loop and $n$ an integer. We will note $I_{SJ} f(\varphi_{SJ})$ the current phase relation of the junction, with $I_{SJ}$ its critical current. In this case the equation for the phase dynamics is~:
\begin{equation}
  		\phi_0 C \frac{d^2\varphi}{dt^2}+\frac{\phi_0}{R} \frac{d\varphi}{dt}+ I_C \sin (\varphi)+I_{SJ} f(\varphi-2\pi\phi/\phi_0)=I
\label{eq:equadiffRCSJ_SQUID}  		
\end{equation}
Similarly to the previous section, this corresponds to the dynamics of a fictitious particle evolving in the potential $U(\varphi,I)$, which is now given by~:
\begin{equation}
	U(\varphi,I)/\phi_0=-I_C \cos (\varphi) + I_{SJ} F(\varphi-2\pi\phi/\phi_0)-I.\varphi,
	\label{potentialSQUID}
\end{equation}
with $F(\varphi)$ a primitive function of the current phase function $f$. This potential is modified by the magnetic flux $\phi$ applied to the SQUID loop. The current phase relation is extracted  from the modulation of the switching current of the SQUID. We will now consider the case of a SQUID with two Josephson junctions and a weak link.
\begin{figure}[htbp]
  \begin{center}
			\includegraphics[width=8cm]{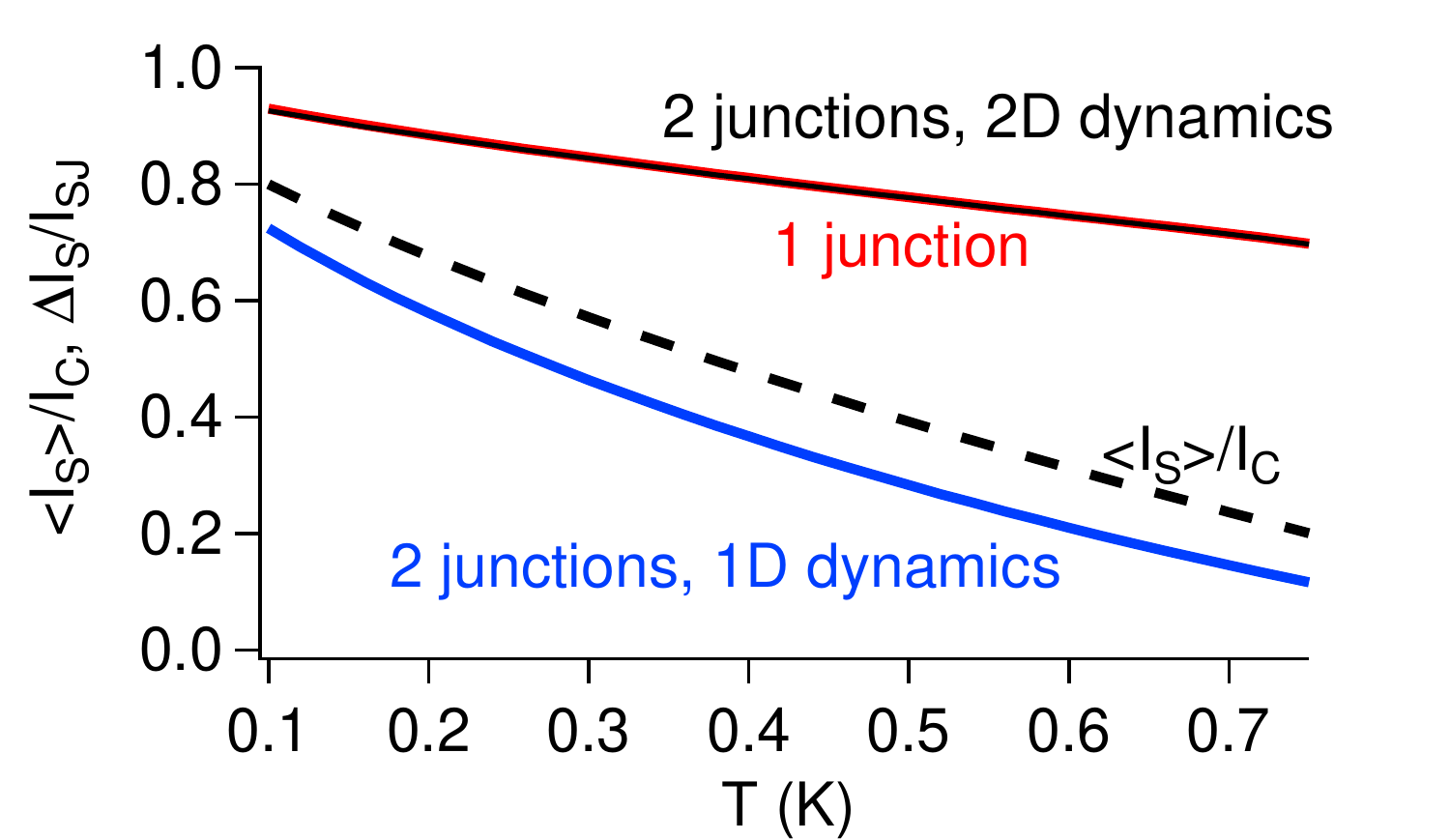}
	\end{center}
  \caption{Calculated $<I_S>$ in unit of $I_C$ (black dashed line) and amplitude of $\Delta I_S$ in unit of the supercurrent $I_{SJ}$ of the weak link for a moderately underdamped SQUID in the thermal escape regime with the parameters $I_C=300$nA, $C_{Tot}=50$fF and $dI/dt=500 \mu$A/s. We have chosen $I_{SJ}=0.05 I_C$ but the calculated ratio $\Delta I_S/I_{SJ}$ is independent of $I_{SJ}$ for $I_{SJ}<0.1 I_C$. Different cases are shown : the case of a SQUID with a single big Josephson junction (red curve) or with two big Josephson junctions with 1D dynamics (black curve) or 2D dynamics (blue curve). Note that the black and red curve are nearly perfectly superimposed. }
  \label{S0_1et2JJ}
\end{figure}

\subsection{Asymmetric SQUID with a central contact}
In the SQUID geometry shown in fig. \ref{fig:RCSJSquid}, we denote $\varphi_1$, $\varphi_2$ the phase difference across the two reference junctions and $\varphi_{SJ}$ the phase across the weak link. Due to fluxo\"id quantization, one now has~:
\begin{equation}
\varphi_{SJ}-(\varphi_1 + \varphi_2) = -2\pi \phi/\phi_0 + 2n\pi
\end{equation}
The critical current of the two big junctions are noted $I_{C1}$ and $I_{C2}$. 

In the following, we consider the three junctions in the framework of the RCSJ model, as shown in figure \ref{fig:RCSJSquid}.
One obtains the following equations relating the currents and the phase difference for each junction~:
\begin{equation}
			i=I_{C1} \sin(\varphi_{1})+\frac{\phi_0}{R_{1}} \frac{d\varphi_1}{dt}+ \phi_0 C_{1}\frac{d^2\varphi_{1}}{dt^2}\\
\end{equation}
where the equation for junction 2 and the weak link can be obtained by replacing "1" by "2" or "$SJ$". In the following we neglect the dynamics of the phase across the small junction compared to the one of the big junctions.  This corresponds to neglecting the small capacitance and conductance of the weak link compared to the ones of the Josephson junction. In this case the previous equations can be recast as :
\begin{eqnarray}
			I=&I_{C1} \sin (\varphi_{1})+I_{SJ}f(\varphi_{SJ})+\frac{\phi_0}{R_{1}} \frac{d\varphi_1}{dt}+ \phi_0 C_{1}\frac{d^2\varphi_{1}}{dt^2} \nonumber \\
			I=&I_{C2} \sin (\varphi_{2})+I_{SJ}f(\varphi_{SJ})+\frac{\phi_0}{R_{2}} \frac{d\varphi_2}{dt}+ \phi_0 C_{2}\frac{d^2\varphi_{2}}{dt^2} \nonumber \\
			\label{eq:eq:equadiffRCSJ_SQUID2D}
\end{eqnarray}
These equations correspond to the dynamics of a fictitious particle evolving in the 2D potential : 
\begin{eqnarray}
&U(\varphi_{1},\varphi_{2},I)/\phi_0=-I_{C1} \cos (\varphi_{1})-I_{C2} \cos (\varphi_{2})  \nonumber \\
&+I_{SJ} F(\varphi_1+\varphi_2-2\pi\phi/\phi_0)-(\varphi_1+\varphi_2).I
\end{eqnarray}
In this situation, depending on the nature of the current noise through the SQUID two regimes can be reached. The first, called hereafter "2D dynamics", corresponds to uncorrelated variations of the phases of each big junction. The second regime, called hereafter "1D dynamics", corresponds to strongly coupled variations of the two phases, which dynamics is synchronized. In this latter situation, since $I_{C1}=I_{C2}=I_0$ and supposing that the two junctions are identical we will make the assumption that, at anytime before the switching, $\varphi_{1}=\varphi_{2}=\varphi$. Then the phase evolves in the effective potential simply given by~: 
\begin{equation}
	U(\varphi,I)=-I_0 \cos (\varphi)+\frac{1}{2} I_{SJ} F(2\varphi-2\pi\phi/\phi_0)-I.\varphi
	\label{eq:potential1D}
\end{equation}
This is very similar to the case of the regular SQUID where the potential is given by Eq. 
\ref{potentialSQUID}.
\begin{figure}[htbp]
  \begin{center}
		\includegraphics[width=8cm]{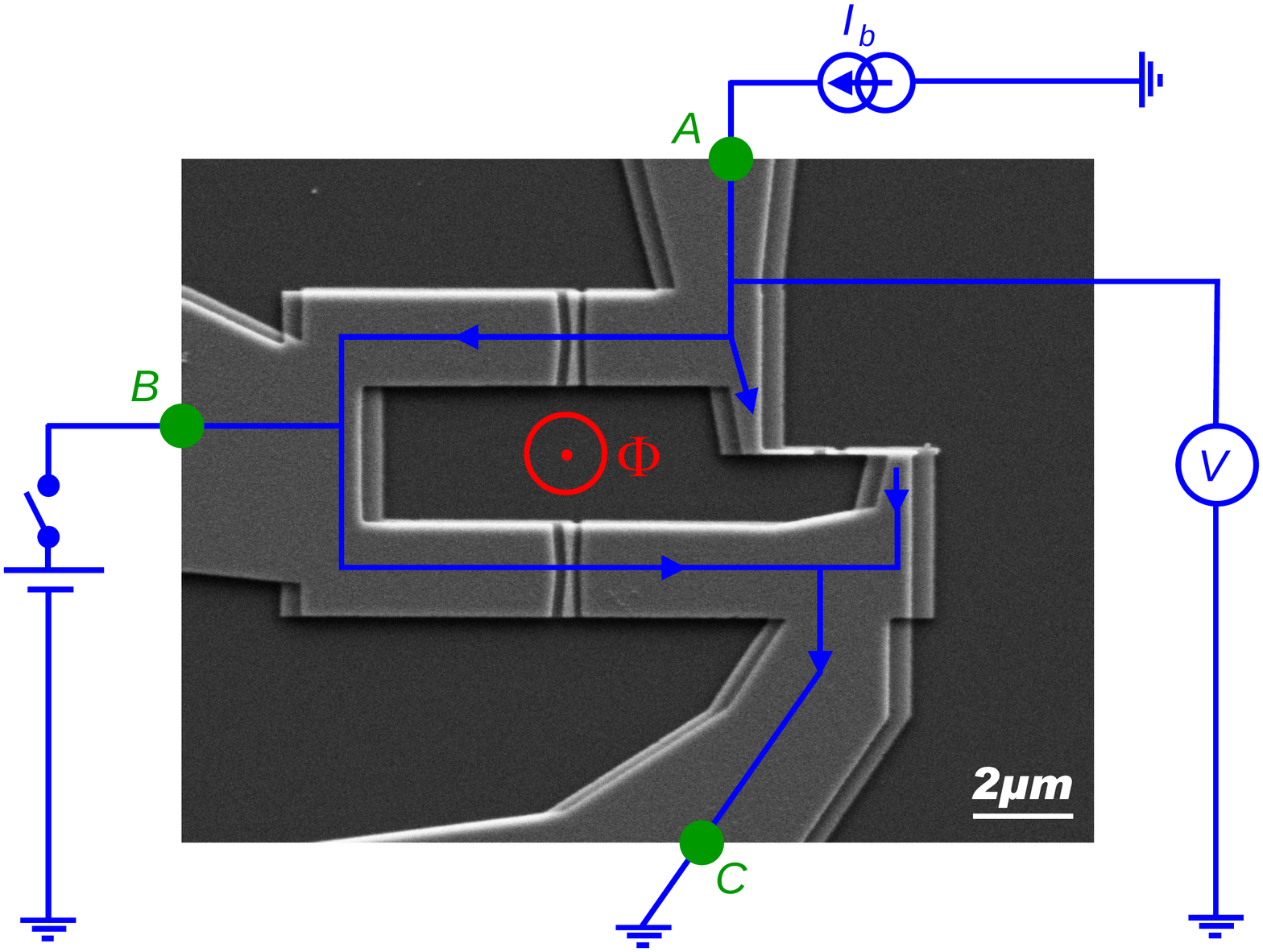} 
		\includegraphics[width=8cm]{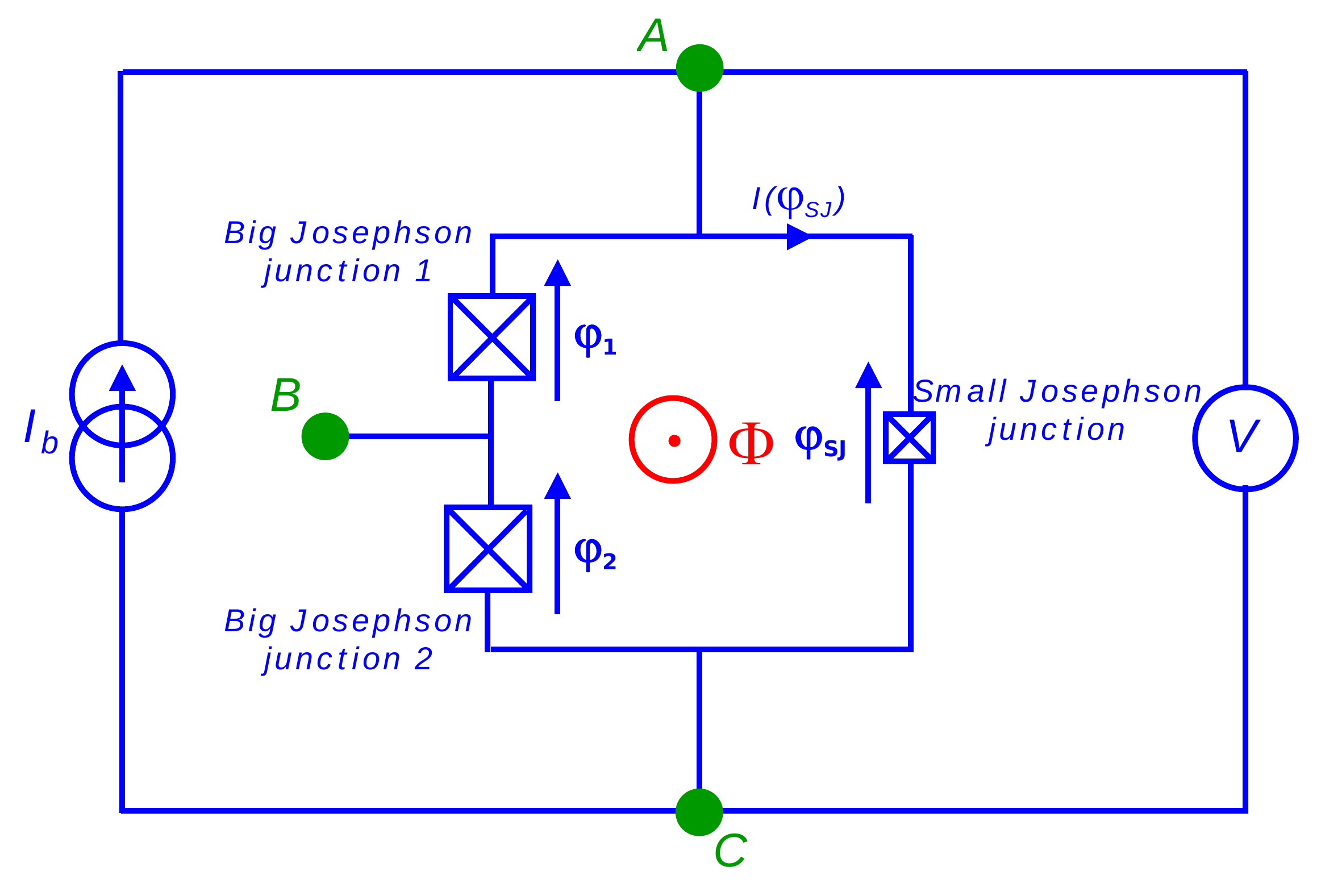}
	\end{center}
  \caption{(a) Scanning electron microscope picture of the asymmetric superconducting SQUID loop. (b) Equivalent circuit of the SQUID. The phase $\varphi_1$, $\varphi_2$ and $\varphi_{SJ}$ are linked to one another by the magnetic flux through the loop $\phi$ by relation $\varphi_{SJ}-(\varphi_1 + \varphi_2) = -2\pi \phi/\phi_0 + 2n\pi$ with $\phi_0 = h/2e$. Contacts $A$ and $C$ are respectively the SQUID source and drain. Contact $B$ is the additional central contact between the two big junctions. It allows to determine the normal state resistance of each junctions.}
  \label{sampleSIS}
\end{figure}

Due to the presence of the central contact, taking into account the current fluctuations in the dynamics of the SQUID is more complicated than in the dc SQUID geometry. This is done phenomenologically by incorporating an effective temperature $T_{esc}$ which can be substantially different from the actual electronic temperature of the experiment but also from the effective temperature in the standard dc SQUID geometry. 

\subsection{Comparison of the two SQUID geometries}
The measurement of the current-phase relation of the weak link is deduced from the modulation of the switching current of the SQUID in the limit where the supercurrent of the Josephson junctions is much higher than the supercurrent of the weak-link. We note the switching current $I_S$, its average value $<I_S>$ and $\Delta I_S(\Phi)$ the magnetic field dependent part, so that: $I_S=<I_S>+\Delta I_S(\Phi)$. In the very low temperature limit $\Delta I_S(\Phi)$ is related to the current phase relation of the weak link and $<I_S>$ is the switching current of the big Josephson junctions. In the next part we address the relation between the measured modulation and the real current phase relation. To do so we have calculated the expected modulation of the moderately underdamped SQUID in the thermal escape regime as a function of temperature, comparing the cases of a two junctions and three junctions SQUID. We suppose that the weak link in one branch of the dc SQUID has a sinusoidal current-phase relation, i.e. $f(\varphi_{SJ})=\sin \varphi_{SJ}$ in eq. \ref{eq:equadiffRCSJ_SQUID} and \ref{eq:eq:equadiffRCSJ_SQUID2D}.

We consider a SQUID submitted to a current bias increasing linearly with time at a rate $dI/dt$. In this case the probability for the system to have switched to the resistive state at a current $I$ is \cite{garg95,ruggiero98,petkovicPhD}~:
\begin{equation*}
	W(I)=1-\exp \left( \int_0^I dI' \frac{\Gamma(I')}{dI/dt}\right)
\end{equation*}
The switching current is determined by solving numerically $W(I)=1/2$ which is equivalent for a moderately underdamped SQUID in the thermal escape regime to:
\begin{equation}
	\int_0^s ds' \omega_P(s') \exp \left( - \frac{\Delta U(s')}{k_B T_{esc}}\right) = \frac{2 \pi}{I_C} \frac{dI}{dt} \ln(2)
	\label{eq:switching}
\end{equation}
with $s=I/I_C$. To calculate the switching current we have taken the following parameters: $I_0=300$nA, $C_{Tot}=50$fF, $dI/dt=500 \mu$A/s and compared the switching current and the amplitude of the modulation (Fig. \ref{S0_1et2JJ}) for the case with one junction and two junctions. 

We see that the situation with a regular SQUID and a SQUID with two big junctions are nearly identical in the 2D dynamics regime. On the other hand the situation in the 1D limit is quite different and leads in particular to a smaller modulation of the SQUID supercurrent. This is related to the reduction by a factor 2 of the effect of the weak-link in the effective potential (equation \ref{eq:potential1D}) and the $2 \varphi$ phase dependence of this term. We thus see that the dynamics of the phase in the SQUID geometry has important consequences regarding the amplitude of the modulation of the supercurrent.

\section{Current-phase relation and conductance measurement of a Josephson junction}
\label{sec:CPR_JJ}

To test the validity of the theoretical analysis of section \ref{sec:CPRsquid} and check the feasibility of the extraction of the current phase relation in a three junctions SQUID configuration we have measured two samples where the weak link consists of a small Josephson junction (Fig. \ref{sampleSIS}). The junctions have been fabricated by electron beam lithography  and shadow evaporation on oxidized Si wafers. For sample 1 the sequence of deposited materials is $Pd(4nm)/Al(70nm)/AlO_X/Al(120nm)$ the same as used for samples with a carbon nanotube (see section \ref{sec:CPR_NT}). Thanks to the three terminal configuration, the resistance of the junctions can be measured~: $R_1=1.48$k$\Omega, R_2= 1.46$k$\Omega$ and  $R_{SJ}=27$k$\Omega$. Given the superconducting gap values $\Delta_{Pd/Al}=160 \mu$eV and $\Delta_{Al}=240 \mu$eV one can calculate the expected critical current $I_{C1}=232$nA, $I_{C2}=235$nA and $I_{SJ}=12.7$nA. Sample 2, that was fabricated with aluminum only ($Al(70nm)/AlO_X/Al(120nm)$), has parameters~: $R_1=2.33k\Omega, R_2= 2.28k\Omega$, $R_{SJ}=18.0k\Omega$ and $\Delta_{Al}=185 \mu$eV, $I_{C1}=125$nA, $I_{C2}=127$nA and $I_{SJ}=16.1$nA.
\begin{figure}[htbp]
	\begin{center}
		\includegraphics[width=8cm]{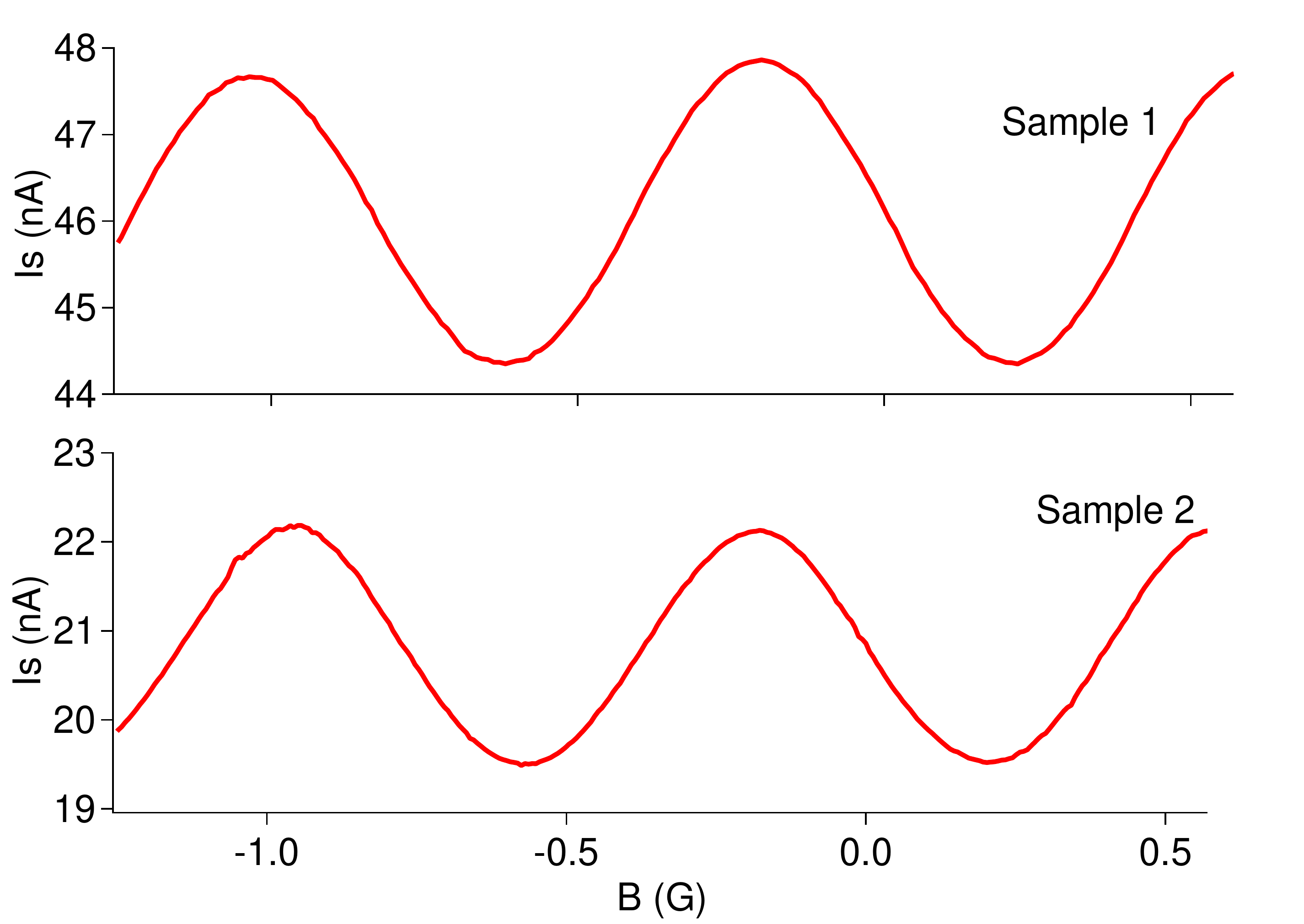}
	\end{center}
	\caption{Modulation of the switching current of the SQUID versus magnetic field for sample 1 (T=230mK) and 2 (T=60mK). In these sample the weak kink of interest is a small superconducting tunnel junction.}
	\label{Is_SIS}
\end{figure}

\subsection{Current Phase relation measurement}
\label{CurPhasMeasSIS}

The switching current of the SQUID is measured by applying a current bias increasing linearly with time and recording the value of the current at which the SQUID switches to a resistive state. This measurement is repeated to obtain an average switching current. This procedure is then repeated for different values of magnetic field. This leads to the magnetic field dependence of the SQUID's switching current (Fig. \ref{Is_SIS} for sample 1 and 2).
\begin{figure}[htbp]
	\begin{center}
		\includegraphics[width=8cm]{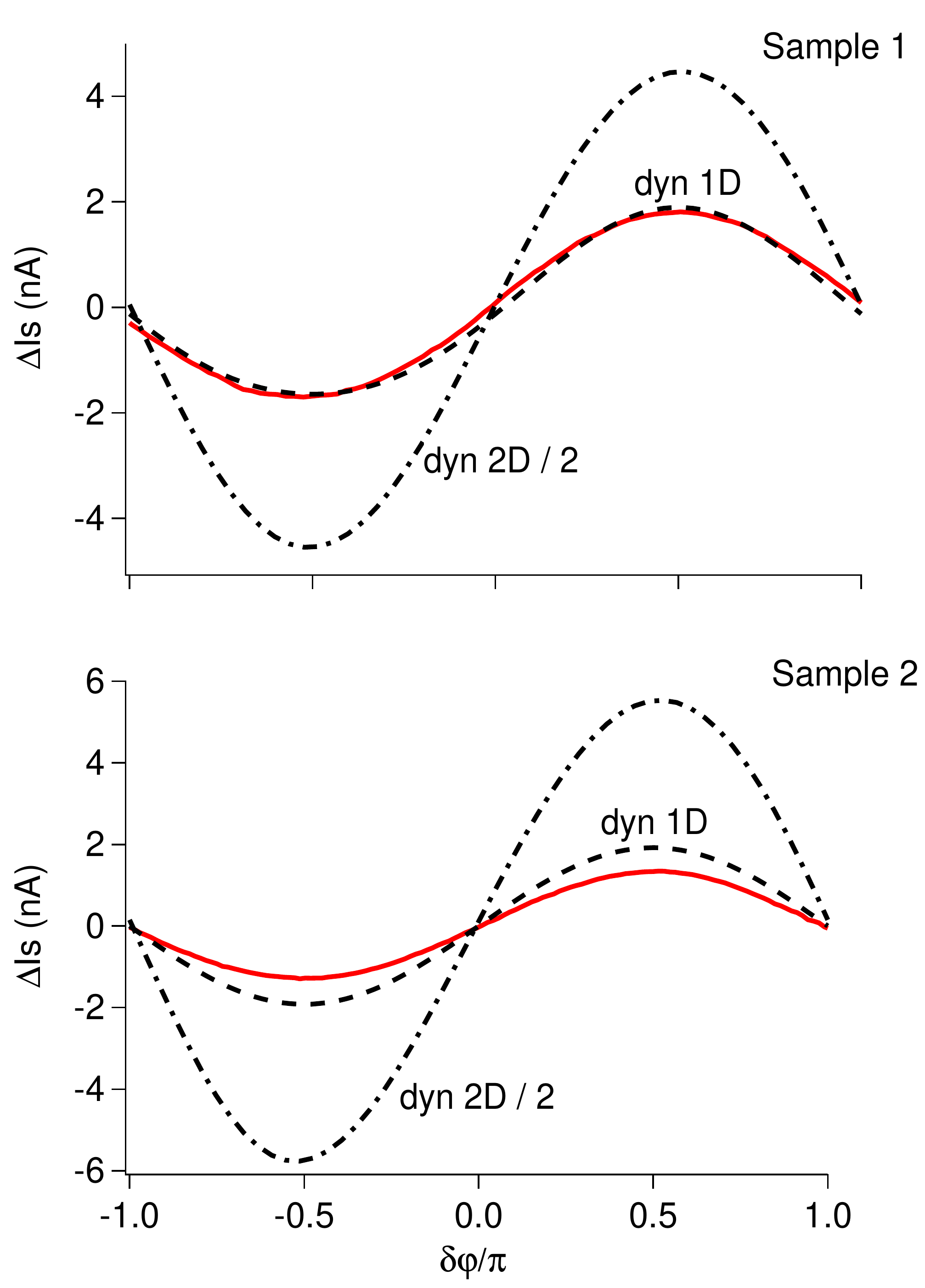}
	\end{center}
	\caption{Modulation of the switching current for sample 1 and 2 (plain red line) and comparison with calculated modulation for the two junction squid with 1D (black dashed line)or 2D dynamics (black dashed-dotted line). Note that the amplitude of the calculated curve in the 2D dynamics is divided by a factor 2 for better comparison.}
	\label{fig:CPR_SIS}
\end{figure} 

For sample 1 the experiment was done at a temperature $T=230$mK. An effective temperature $T_{esc}=460$mK is needed to explain the low switching current $I_S=46$nA compared to the critical current $I_C=234$nA (see relation \ref{eq:switching}). As noted before since it is more difficult to take into account the current noise in the two reference junctions squid geometry, this effective temperature can be substantially different from the case of the standard DC squid geometry. This unusually high value of phase temperature may also indicate the fact that our experiment is not in the purely thermally activated regime but can also exhibit phase diffusion \cite{vion96}. With this effective temperature we calculate the expected modulation of the switching current with a 1D or 2D dynamics to define which model is most relevant for our experiment (Fig. \ref{fig:CPR_SIS}). The calculated modulation in the 2D case is nearly 5 times bigger than the measured one, whereas the 1D case gives a good agreement with the experimental data. For sample 2, the 2D calculation overestimates by a factor 6 the measured switching current modulation. The 1D calculation is in agreement with the experiment within 30 \%.
\begin{figure}[htbp]
  \begin{center}
		\includegraphics[width=8cm]{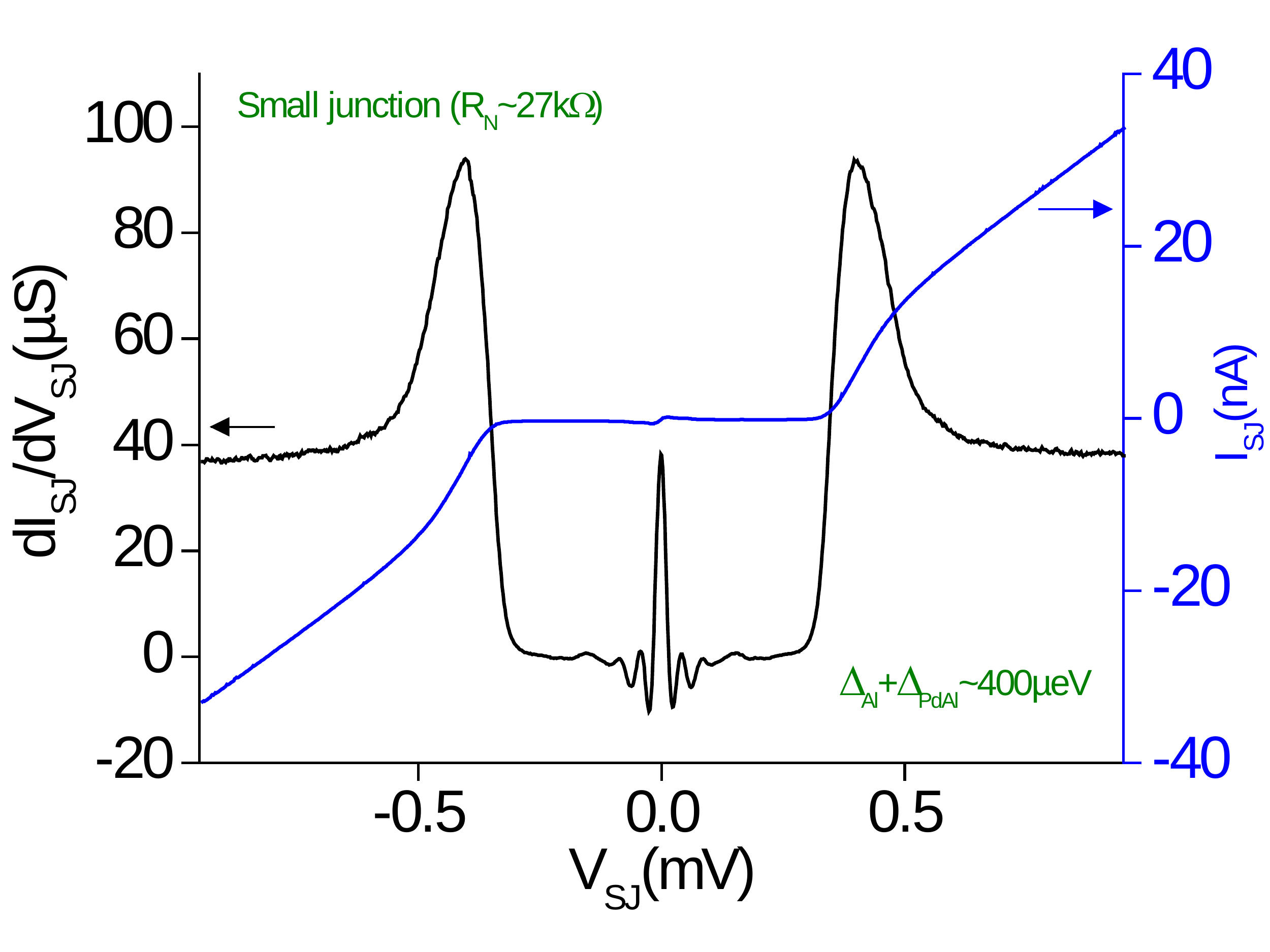}
	\end{center}
  \caption{Differential conductance $dI_{SJ}/dV_{SJ}$ and dc characteristics $I_{SJ}(V_{SJ})$  of the small junction. The traces are obtained by fixing $V_{BC}$ below the superconducting gap of the corresponding  large junction. Superconducting gap is found to be $\Delta_{PdAl} + \Delta_{Al}=400\mu eV$ and the normal state resistance $R_{N,SJ}=27k\Omega$.}
  \label{dIdVSIS}
\end{figure} 

The sinusoidal shape of the current phase relation, expected for this Josephson junction, is found correctly in the experiment. However the quantitative agreement with the presented theory is not completely satisfactory.  This result may motivate more involved calculations in this new three terminal SQUID structure. 

\subsection{$dI/dV$ of the small junction in the superconducting state}
\label{CondSupraSquidSIS}

The differential conductance $dI/dV$ in the superconducting state was measured by applying a dc voltage bias on the SQUID (between $A$ and $C$) and monitoring the current flowing out of point $C$ while maintaining the bias voltage at point B such that $V_{BC}$ stays below the superconducting gap of the junction.  By doing so it was possible to extract the differential conductance of the small Josephson junction with this scheme (Fig. \ref{dIdVSIS}) provided that the Josephson branch of the big Josephson junctions are suppressed by a magnetic flux equal to a flux quantum in the area of the big junctions.
\begin{figure}[htbp]
	\begin{center}
		\includegraphics[width=8cm]{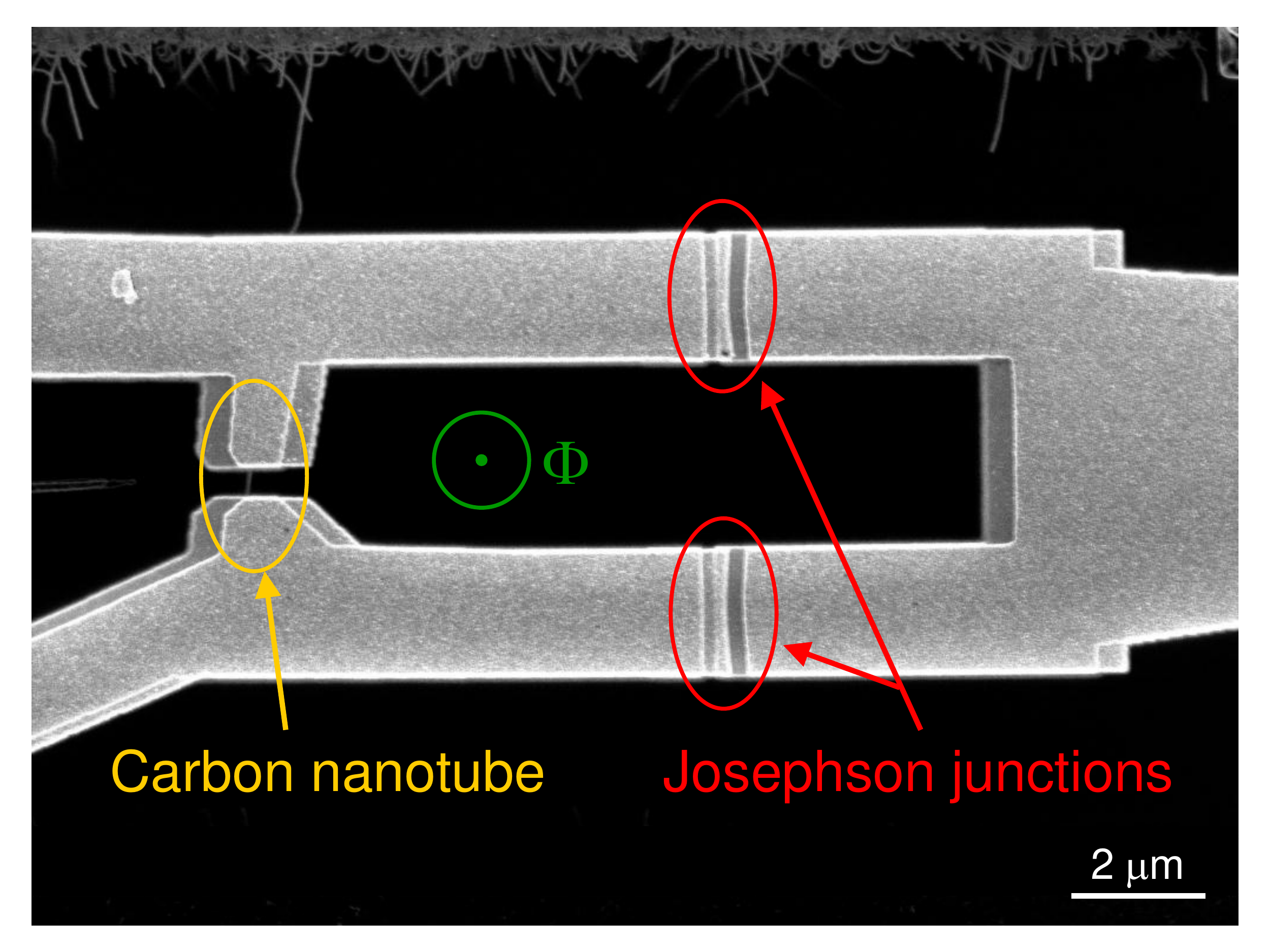}
	\end{center}
	\caption{Scanning electron microscope picture of the asymmetric SQUID used in the experiment. Junctions and carbon nanotube contacts are made of $PdAl/Al0_X/Al$ and are fabricated in the same step of metal deposition. Nanotube contacts are $450nm$ apart.}
	\label{PhotoCourPhaseTube}
\end{figure} 

\subsection{Conclusion}
In this last section we have demonstrated the possibility to measure on the same sample the current phase relation and the differential conductance in the superconducting state by using a SQUID geometry with two reference Josephson junctions and three contacts. Hereafter we use the same detection scheme to measure the current phase relation of a carbon nanotube quantum dot strongly coupled to superconducting leads. 

\section{Current Phase relation of a carbon nanotube quantum dot junction}
\label{sec:CPR_NT}

\subsection{Sample Fabrication}

The design of the sample is similar to sample 1 described previously except that the weak link is now constituted by a carbon nanotube junction. The carbon nanotube is grown by chemical vapor deposition \cite{kasumov07} and is connected with $PdAl/Al0_X/Al$ contacts in the same run of deposition as the big junctions in parallel (Fig. \ref{PhotoCourPhaseTube}). The three points measurements at room temperature allows to determine the resistance of each junction: $R_1=1.03k\Omega, R_2= 1.02k\Omega$.
\begin{figure}[htbp]
	\begin{center}
		\includegraphics[width=8cm]{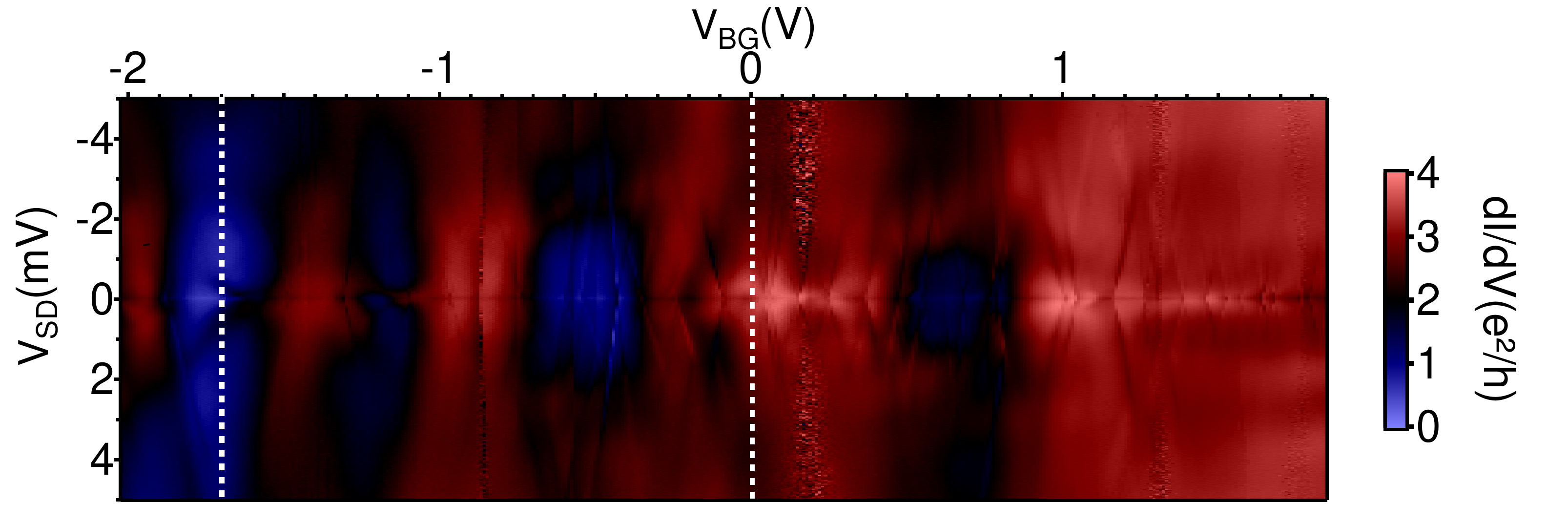}
	\end{center}
	\caption{Normal state stability diagram of the carbon nanotube quantum dot \textit{i.e.} differential conductance $dIdV$ \textit{vs} bias voltage $V_{SD}$ and back-gate voltage $V_{BG}$.}
	\label{fig:NTnormal}
\end{figure}

\subsection{Normal state characterization of the carbon nanotube quantum dot}

The normal state characterization of the carbon nanotube at low temperature is achieved by first applying a magnetic field $B\approx 0.18T$ which suppresses superconductivity in the contacts. The differential conductance $dI/dV$ is then measured with a lock-in amplifier as a function of bias and back-gate voltages (Fig. \ref{fig:NTnormal}). The nanotube is globally highly conducting with a maximum differential conductance approaching $4 e^2/h$.
\begin{figure}[htbp]
	\begin{center}
		\includegraphics[width=8cm]{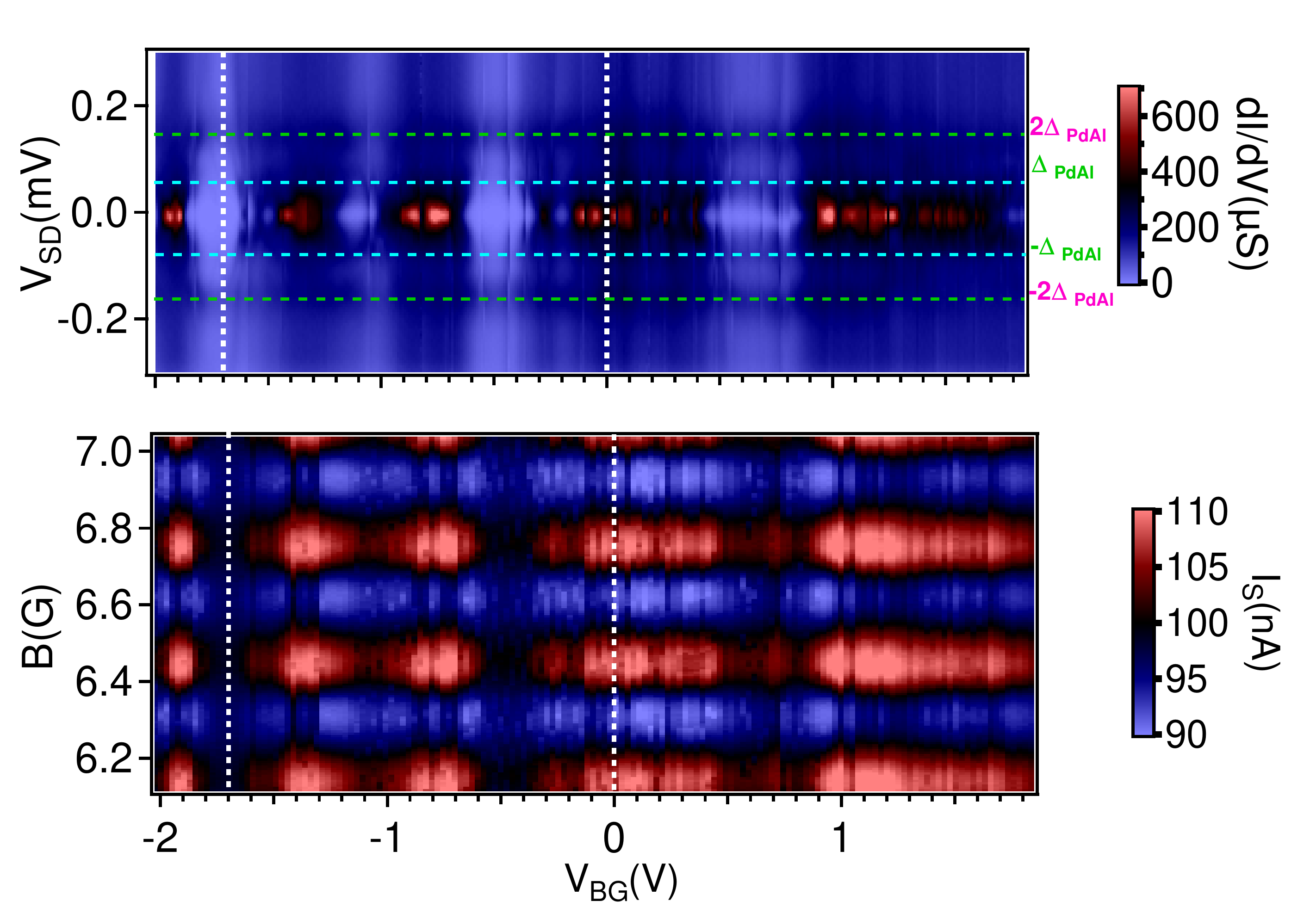}
	\end{center}
	\caption{Top panel : differential conductance dI/dV of the carbon nanotube quantum dot in the superconducting state. The two vertical dashed lines indicate cuts at a given backgate voltage shown in figure \ref{fig:Nanotube_Ic_dIdVs}. The horizontal dashed lines indicate the position of the multiple Andreev reflections. Bottom panel : Modulation of the SQUID supercurrent \textit{vs} the applied magnetic field $B$ and the back-gate voltage.}
	\label{fig:NTsupra}
\end{figure}

\subsection{Superconducting state characterization of the carbon nanotube quantum dot}
To measure $dI/dV$ in the superconducting state, we reduce the magnetic field below the critical field of the contact and used the technique described in section \ref{sec:CPR_JJ} (Fig. \ref{fig:NTsupra}). Two traces are also shown in fig \ref{fig:Nanotube_Ic_dIdVs}a. We observe zero bias peaks in the regions of high normal state conductances and dips in the low conductance region. In addition to this, multiple Andreev reflections are also visible at fixed voltages $2\Delta/n$ with $n$ an integer number ($n=\pm1,\pm2$ here). Finally, far from the superconducting gap ($|V_{SD}|>>2\Delta$), the conductance is constant.  

\subsection{Current phase relation measurement}
\begin{figure}
	\begin{center}
		\includegraphics[width=8cm]{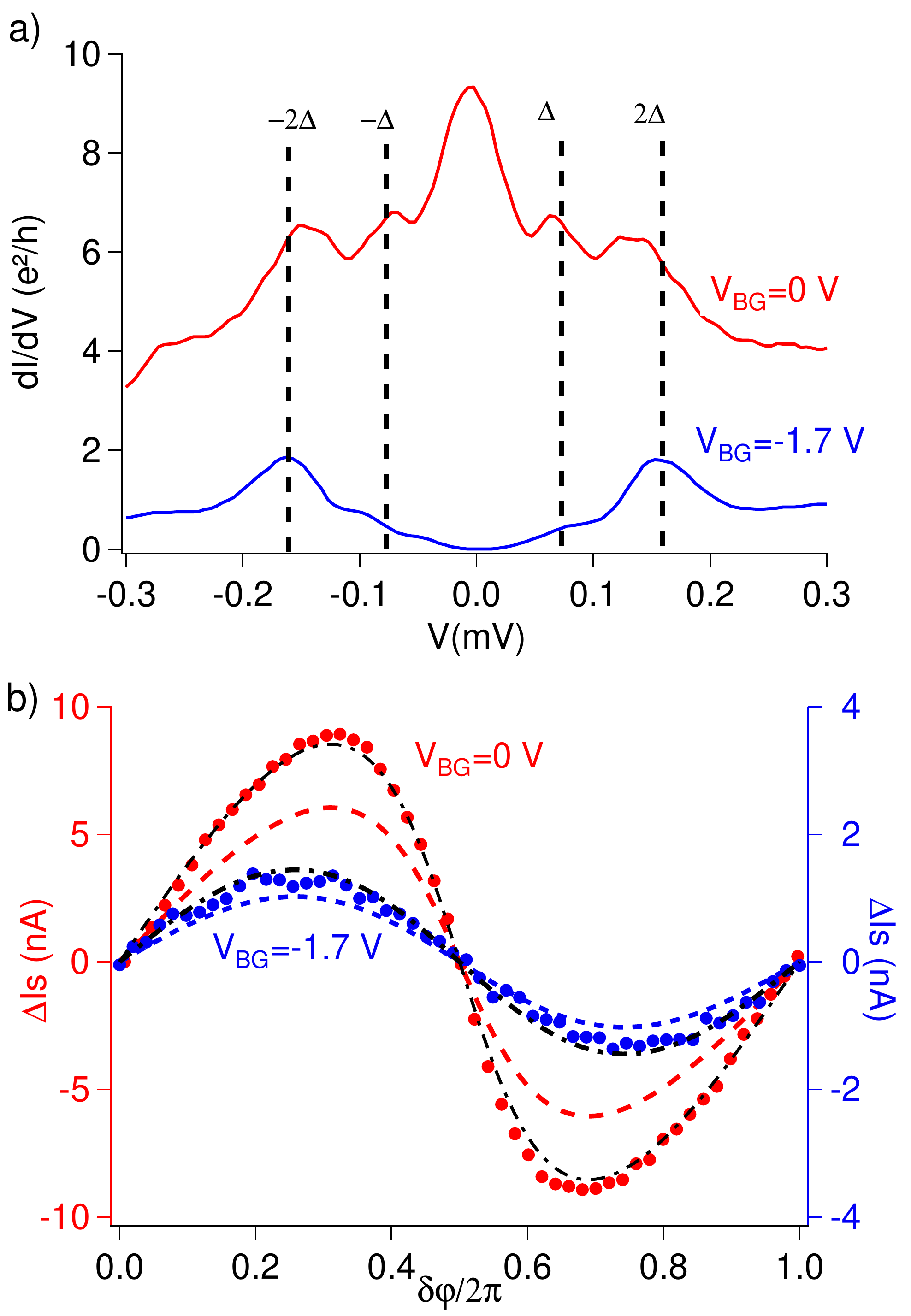}
	\end{center}
	\caption{(a) Differential conductance of the nanotube junction in the superconducting state at two gate voltages, $V_G=0V$ (in red) corresponding to a high normal state conductance and $V_G=-1.7V$ (in blue) corresponding to a low conductance. The two curves show evidence of multiple Andreev reflection.(b) Current phase relation extracted from the experiment at same gate voltages (red and blue circles). The dashed lines are theoretical predictions based on eq. \ref{eq:IphiTrans} and the 1D dynamics model. The dashed dotted lines correspond to eq.\ref{eq:IphiTrans} scaled by a factor 0.36.} 
	\label{fig:Nanotube_Ic_dIdVs}
\end{figure}

In the superconducting state the SQUID exhibits a modulation of its supercurrent versus magnetic flux over the entire investigated range of gate voltage, a proof that the nanotube junction carries a supercurrent over that gate voltage range. In order to perform a quantitative analysis we focus on two gate voltages, one corresponding to a high conductance in the normal state of the junction ($V_G=0$ V) and one which is less conducting ($V_G=-1.7$ V). For each gate voltage we have measured both the current phase relation extracted from the modulation of the switching current (Fig. \ref{fig:Nanotube_Ic_dIdVs}b) and the differential conductance $dI/dV$ in the superconducting state (Fig. \ref{fig:Nanotube_Ic_dIdVs}a). At some gate voltages supercurrents as high as $12$ nA were induced through the tube and the current-phase relation exhibits an anharmonic behaviour.

We relate the shape and amplitude of the current-phase relation to the normal state conductance at zero bias. We measured for $V_G=0$V, $G=3.19 e^2/h$ and $G=0.7 e^2/h$ for $V_G=-1.7$V. Considering the carbon nanotube as a conductor with two spin degenerate conducting channels with the same transmission $\tau$ one gets $\tau=0.79$ at $V_G=0$V and $\tau=0.175$ at $V_G=-1.7$V. Such channels, with transmission $\tau_i$ between two superconducting contacts of gap $\Delta$ are expected to show a current-phase relation given by \cite{beenaker91,rodero94}~: 
\begin{equation}
	I(\varphi)=\sum_i \frac{e \tau_i \Delta}{2 \hbar} \frac{\sin(\varphi)}{\sqrt{1-\tau_i \sin^2(\varphi/2)}}
	\label{eq:IphiTrans}
\end{equation}

The experiment was done at a temperature $T=30$mK. To explain the value of the switching current $I_S=100$nA compared to the value of the critical current $I_C=335$nA we have to take in relation \ref{eq:switching} an effective temperature $T_{esc}=597mK$. With this effective temperature we calculate a reduction factor for the amplitude of the switching current modulation of 0.765 for the 2D dynamics and 0.255 for the 1D dynamics. As noted in the previous section the agreement with the measured amplitude is best with the 1D dynamics model and is within 30 \% for the two gate values shown. To obtain a better agreement a deeper understanding of the switching in our device is needed. The amplitude of the modulation of the switching current $\Delta I_S$ compared to the theoretical value deduced from eq. \ref {eq:IphiTrans} is found to be 0.36 (black dashed line in fig. \ref{fig:Nanotube_Ic_dIdVs}b). We want to stress that the shape of the expected current-phase relation is well reproduced in the experiment. 

\section{Conclusion}

Our detection setup allows to relate the current phase relation measurements to the normal and superconducting states differential conductance $dI/dV$. This provides a useful way to measure precisely the current phase relation and parameters of more complex system. It might in particular be extremely interesting and challenging to probe the signature of electronic correlation in conjunction with proximity effect or the influence of large spin-orbit interactions. 

\section{Acknowledgements}

We acknowledge M. Aprili, J. Gabelli, C. Ojeda-Aristizabal, M. Ferrier and S. Gu{\'e}ron for fruitful discussions. This work has benefited from financial support of ANR under project MASH (ANR-12-BS04-0016-MASH).

\end{document}